\documentstyle[epsf,12pt]{article}

\def\baselinestretch{1.5}
\newcommand{\be}{\begin{equation}}
\newcommand{\ee}{\end{equation}}

\begin{document}

\title{{\bf Multifractal characterisation of length sequences of coding and
noncoding segments in a complete genome}}
\author{Zu-Guo Yu$^{1,2}$\thanks{%
Corresponding author, corresponding address: School of Mathematical Science,
Queensland University of Technology, Garden Point Campus, GPO Box 2434,
Brisbane, Q 4001, Australia. Tel.: +61 7 38645194, Fax: +61 7 38642310,
e-mail: yuzg@hotmail.com or z.yu@qut.edu.au}, Vo Anh$%
^{1}$ and Ka-Sing Lau$^3$ \\
{\small $^1$Centre in Statistical Science and Industrial Mathematics,
Queensland University} \\
{\small of Technology, GPO Box 2434, Brisbane, Q 4001, Australia.}\\
{\small $^2$Department of Mathematics, Xiangtan University, Hunan 411105, P.
R. China.\thanks{
This is the permanent corresponding address of Zu-Guo Yu.}}\\
{\small $^3$Department of Mathematics, Chinese University of Hong Kong,
Shatin, Hong Kong}}
\date{24 April 2001}
\maketitle

\begin{abstract}
The coding and noncoding length sequences constructed from a complete genome
are characterised by multifractal analysis. The dimension spectrum $D_{q}$
and its derivative, the 'analogous' specific heat $C_{q}$, are calculated
for the coding and noncoding length sequences of bacteria, where $q$ is the
moment order of the partition sum of the sequences. From the shape of the $%
D_{q}$ and $C_{q}$ curves, it is seen that there exists a clear difference
between the coding/noncoding length sequences of all organisms considered
and a completely random sequence. The complexity of noncoding length
sequences is higher than that of coding length sequences for bacteria.
Almost all $D_{q}$ curves for coding length sequences are flat, so their
multifractality is small whereas almost all $D_{q}$ curves for noncoding
length sequences are multifractal-like. It is seen that the 'analogous'
specific heats of noncoding length sequences of bacteria have a rich variety
of behaviour which is much more complex than that of coding length
sequences. We propose to characterise the bacteria according to the types of
the $C_{q}$ curves of their noncoding length sequences. This new type of
classification allows a better understanding of the relationship among
bacteria at the global gene level instead of nucleotide sequence level.
\end{abstract}

\vskip 0.2cm



\section{Introduction}

\ \ \ \ The rapidly accumulating complete genome sequences of bacteria and
archaea provide a new type of information resource for understanding gene
functions and evolution$^{\cite{gelfand00}}$.

One can study the DNA sequences with details by considering the order of
four kinds of nucleotides which DNA is assembled, namely adenine ($a$),
cytosine ($c$), guanine ($g$), and thymine ($t$).

There has been considerable interest in the finding of long-range
correlation (LRC) in DNA sequences at this level. Li {\it et al}$^{\cite{li}%
} $ found that the spectral density of a DNA sequence containing mostly
introns shows $1/f^{\beta }$ behaviour, which indicates the presence of LRC.
The correlation properties of coding and noncoding DNA sequences were also
studied by Peng {\it et al}$^{\cite{peng}}$ in their fractal landscape or
DNA walk model. The DNA walk defined in Ref.\cite{peng} is that the walker
steps ``up'' if a pyrimidine ($c$ or $t$) occurs at position $i$ along the
DNA chain, while the walker steps ``down'' if a purine ($a$ or $g$) occurs
at position $i$. Peng {\it et al}$^{\cite{peng}}$ discovered that there
exists LRC in noncoding DNA sequences while the coding sequences correspond
to a regular random walk. By doing a more detailed analysis,
Chatzidimitriou-Dreismann and Larhammar$^{\cite{CDL93}}$ concluded that both
coding and noncoding sequences exhibit LRC. A subsequent work by Prabhu and
Claverie$^{\cite{PC92}}$ also substantially corroborates these results. If
one considers more details by distinguishing $c$ from $t$ in pyrimidine, and 
$a$ from $g$ in purine (such as two or three-dimensional DNA walk model$^{
\cite{luo}}$ and maps given in Ref.\cite{YC}), then the presence of base
correlation has been found even in coding sequences. In view of the
controversy about the presence of correlation in all DNA or only in
noncoding DNA, Buldyrev {\it et al}$^{\cite{Bul95}}$ showed the LRC appears
mainly in noncoding DNA using all the DNA sequences available.
Alternatively, Voss$^{\cite{voss}}$, based on equal-symbol correlation,
showed a power law behaviour for the sequences studied regardless of the
percent of intron contents. Investigations based on different models seem to
suggest different results, as they all look into only a certain aspect of
the entire DNA sequence$^{\cite{MNR}}$.

The avoided and under-represented strings in some bacterial complete genomes
have been discussed$^{\cite{yhxc99,hlz98,hxyc99}}$. A time series model of
CDS in complete genome has been proposed$^{\cite{YW99}}$. Vieira$^{\cite
{Vie99}}$ performed a low-frequency analysis of the complete DNA of 13
microbial genomes and found that their fractal behaviour does not always
prevail through the entire chain and their autocorrelation functions have a
rich variety of behaviours including the presence of anti-persistence.

For the importance of the numbers, sizes and ordering of genes along the
chromosome, one can refer to Part 5 of Lewin$^{\cite{Lew97}}$. Here one may
ignore the composition of the four kinds of bases in coding and noncoding
segments and only considers the rough structure of the complete genome or
long DNA sequences. Provata and Almirantis $^{\cite{PY}}$ proposed a fractal
Cantor pattern of DNA. They map coding segments to filled regions and
noncoding segments to empty regions of random Cantor set and then calculate
the fractal dimension of the random fractal set. They found that the
coding/noncoding partition in DNA sequences of lower organisms is
homogeneous-like, while in the higher eucariotes the partition is fractal.
This result seems too rough to distinguish bacteria because the fractal
dimensions of bacteria they gave out are all the same.

Viewing from the level of structure, the complete genome of an organism is
made up of coding and noncoding segments. Here the length of a
coding/noncoding segment means the number of its bases. Based on the lengths
of coding/noncoding segments in the complete genome, one can get two kinds
of integer sequences by the following ways:

i) Order all lengths of coding segments according to the order of coding
segments in the complete genome. This integer sequence is named {\it coding
length sequence}.

ii) Order all lengths of noncoding segments according to the order of
noncoding segments in the complete genome. This integer sequence is named 
{\it noncoding length sequence}.

Yu and Anh$^{\cite{YA00}}$ proposed a time series model for the length
sequences of DNA. After calculating the correlation dimensions and Hurst
exponents, it was found that one can get more information from this model
than that of fractal Cantor pattern$^{\cite{PY}}$. The quantification of
these correlations could give an insight into the role of the ordering of
genes on the chromosome. Through detrended fluctuation analysis (DFA)$^{\cite
{GPH}}$ and spectral analysis, the LRC was found in these length sequences$^{
\cite{YAW01}}$.

The correlation dimension and Hurst exponent are parameters of global
analysis. Global calculations neglect the fact that length sequences from a
complete genome are highly inhomogeneous. Thus multifractal analysis is a
useful way to characterise the spatial inhomogeneity of both theoretical and
experimental fractal patterns$^{\cite{GP83}}$. It was initially proposed to
treat turbulence data. In recent years it has been applied successfully in
many different fields including time series analysis$^{\cite{pas97}}$ and
financial modelling$^{\cite{can00,ATT00}}$. For DNA sequences, application
of the multifractal technique seems rare (we have found only Berthelsen {\it %
et al.}$^{\cite{BGR94}})$. Recently, Yu {\it et al}$^{\cite{YA00a}}$ considered
the multifractal property of the measure representation of a complete
genome. In this paper, we pay more attention to the multifractal
characterisation of the coding and noncoding length sequences.

Some sets of physical interest have a nonanalytic dependence of dimension
spectrum $D_{q}$ on the $q$-moments of the partition sum of the sequences.
Moreover, multifractality has a direct analogy to the phenomenon of phase
transition in condensed-matter physics$^{\cite{KP87}}$. The existence and
type of phase transitions might turn out to be a worthwhile characterisation
of universality classes for the structures$^{\cite{Boj87}}$. The concept of
phase transition in multifractal spectra was introduced in the study of
logistic maps, Julia sets and other simple systems. Evidence of phase
transition was found in the multifractal spectrum of diffusion-limited
aggregation$^{\cite{LeS88}}$. By following the thermodynamic formulation of
multifractal measures, where $q$ represents an analogous temperature, Canessa%
$^{\cite{can00}}$ applied a standard expression for the 'analogous' specific
heat and showed that its form resembles a classical phase transition at a
critical point for financial time series.

In this paper we calculate the 'analogous' specific heat of coding and
noncoding length sequences. Our motivation to apply Canessa's framework to
characterise stochastic sequences is to see whether there is a similar type
of phase transition in the coding and noncoding length sequences as in other
time series. We show that based on the shape of the $C_{q}$ curves and
associated type of phase transitions, one can discuss the classification of
bacteria. This new type of classification allows to better understand the
relationship among bacteria at the global gene level instead of nucleotide
sequence level.

\section{Multifractal analysis}

\ \ \ \ Let $T_{t},\ t=1,2,\cdots ,N,$ be the length sequence of coding or
noncoding segments in the complete genome of an organism. First we define 
\begin{equation}
F_{t}=T_{t}/(\sum_{j=1}^{N}T_{j})
\end{equation}
to be the frequency of $T_{t}$. It follows that $\sum_{t}F_{t}=1$. Now we
can define a measure $\mu $ on $[0,1[$ by $d\mu (x)=Y(x)dx$, where 
\begin{equation}
Y(x)=N\times F_{t},\ \ \mbox{when}\ \ x\in \lbrack \frac{t-1}{N},\frac{t}{N}%
[.
\end{equation}
It is easy to see that $\int_{0}^{1}d\mu (x)=1$ and $\mu
([(t-1)/N,t/N[)=F_{t}$.

The most common numerical implementations of multifractal analysis are the
so-called {\it fixed-size box-counting algorithms} $^{\cite{hjkps}}$. In the
one-dimensional case, for a given measure $\mu $ with support $E\subset {\bf %
R}$, we consider the {\it partition sum} 
\begin{equation}
Z_{\epsilon }(q)=\sum_{\mu (B)\neq 0}[\mu (B)]^{q},
\end{equation}
$q\in {\bf R}$, where the sum runs over all different nonempty boxes $B$ of
a given side $\epsilon $ in a grid covering of the support $E$, that is, 
\begin{equation}
B=[k\epsilon ,(k+1)\epsilon \lbrack .
\end{equation}
The scaling exponent $\tau (q)$ is defined by 
\begin{equation}
\tau (q)=\lim_{\epsilon \rightarrow 0}\frac{\log Z_{\epsilon }(q)}{\log
\epsilon }
\end{equation}
and the generalized fractal dimensions of the measure are defined as 
\begin{equation}
D_{q}=\tau (q)/(q-1),\ \ \mbox{for}\ q\neq 1,
\end{equation}
and 
\begin{equation}
D_{q}=\lim_{\epsilon \rightarrow 0}\frac{Z_{1,\epsilon }}{\log \epsilon },\
\ \mbox{for}\ q=1,
\end{equation}
where $Z_{1,\epsilon }=\sum_{\mu (B)\neq 0}\mu (B)\log \mu (B)$. The
generalized fractal dimensions are numerically estimated through a linear
regression of 
\[
\frac{1}{q-1}\log Z_{\epsilon }(q)
\]
against $\log \epsilon $ for $q\neq 1$, and similarly through a linear
regression of $Z_{1,\epsilon }$ against $\log \epsilon $ for $q=1$. $D_{1}$
is called the {\it information dimension} and $D_{2}$ the {\it correlation
dimension}. The $D_{q}$ of the positive values of $q$ give relevance to the
regions where the measure is large, i.e., to the coding or noncoding
segments which are relatively long. The $D_{q}$ of the negative values of $q$
deal with the structure and the properties of the most rarefied regions of
the measure, i.e. to the segments which are relatively short.

By following the thermodynamic formulation of multifractal measures, Canessa$%
^{\cite{can00}}$ derived an expression for the 'analogous' specific heat as 
\begin{equation}
C_{q}\equiv -\frac{\partial ^{2}\tau (q)}{\partial q^{2}}\approx 2\tau
(q)-\tau (q+1)-\tau (q-1).
\end{equation}
He showed that the form of $C_{q}$ resembles a classical phase transition at
a critical point for financial time series. In the following we calculate
the 'analogous' specific heat of coding and noncoding length sequences for
the first time. The types of phase transitions are helpful to discuss the
classification of bacteria.

\section{ Data and results}

\ \ \ \ More than 31 bacterial complete genomes are now available in public
databases. There are five Archaebacteria: {\it Archaeoglobus fulgidus}
(aful), {\it Pyrococcus abyssi} (pabyssi), {\it Methanococcus jannaschii}
(mjan), {\it Aeropyrum pernix} (aero) and {\it Methanobacterium
thermoautotrophicum} (mthe); five Gram-positive Eubacteria: {\it %
Mycobacterium tuberculosis} (mtub), {\it Mycoplasma pneumoniae} (mpneu), 
{\it Mycoplasma genitalium} (mgen), {\it Ureaplasma urealyticum} (uure), and 
{\it Bacillus subtilis} (bsub). The others are Gram-negative Eubacteria,
which consist of two Hyperthermophilic bacteria: {\it Aquifex aeolicus}
(aquae) and {\it Thermotoga maritima} (tmar); three Chlamydia: {\it %
Chlamydia trachomatisserovar} (ctra), {\it Chlamydia muridarum} (ctraM), and 
{\it Chlamydia pneumoniae} (cpneu); two Spirochaete: {\it Borrelia
burgdorferi} (bbur) and {\it Treponema pallidum} (tpal); one Cyanobacterium: 
{\it Synechocystis sp. PCC6803} (synecho); and thirteen Proteobacteria. The
thirteen Proteobacteria are divided into four subdivisions, which are alpha
subdivision: {\it Rhizobium sp. NGR234} (pNGR234) and {\it Rickettsia
prowazekii} (rpxx); gamma subdivision: {\it Escherichia coli} (ecoli), {\it %
Haemophilus influenzae} (hinf), {\it Xylella fastidiosa} (xfas), {\it Vibrio
cholerae} (vcho1), {\it Pseudomonas aeruginosa} (paer) and {\it Buchnera sp.
APS} (buch); beta subdivision: {\it Neisseria meningitidis MC58} (nmen) and 
{\it Neisseria meningitidis Z2491} (nmenA); epsilon subdivision: {\it %
Helicobacter pylori J99} (hpyl99), {\it Helicobacter pylori 26695} (hpyl)
and {\it Campylobacter jejuni} (cjej).

\begin{figure}[tbp]
\centerline{\epsfxsize=8cm \epsfbox{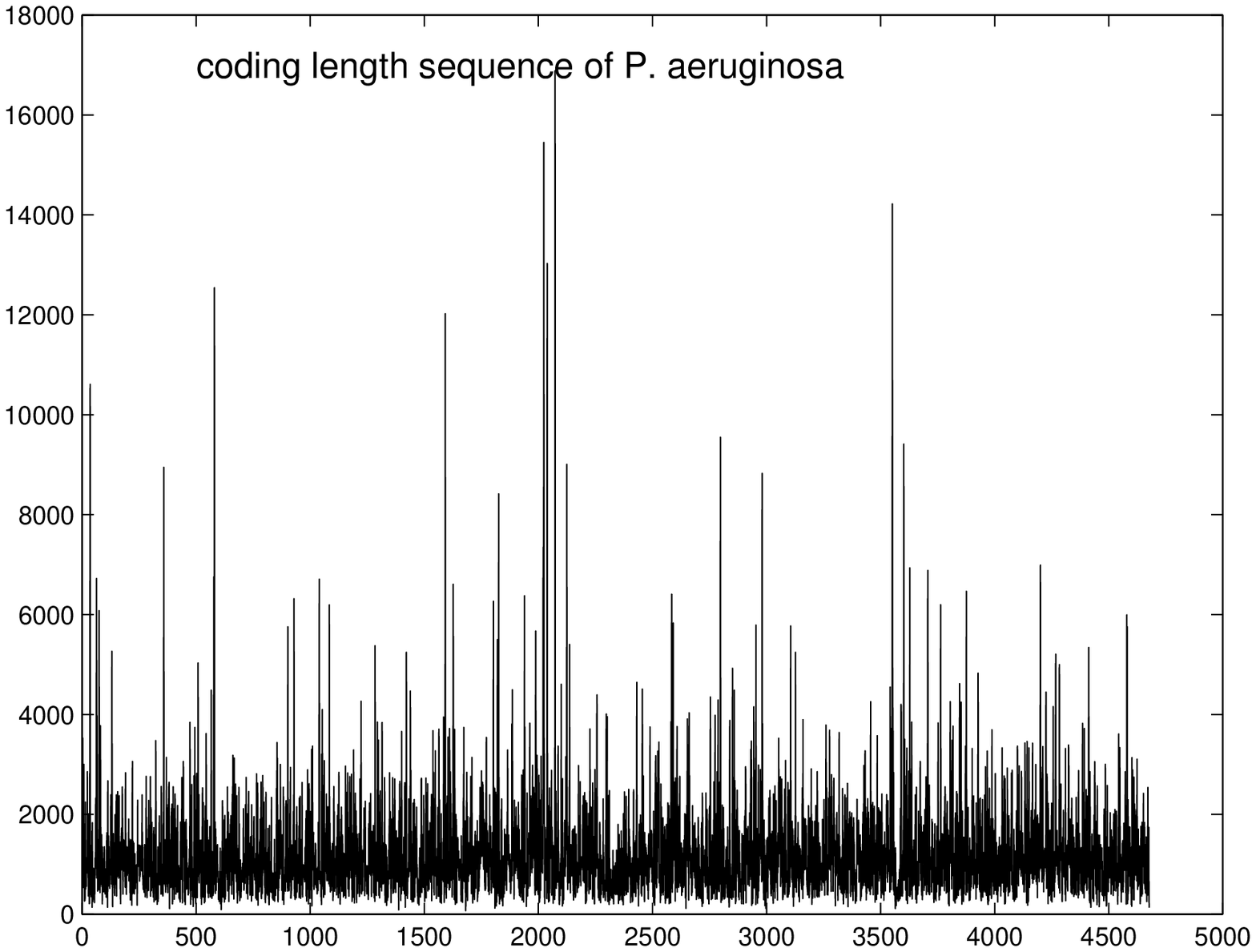}
\epsfxsize=8cm \epsfbox{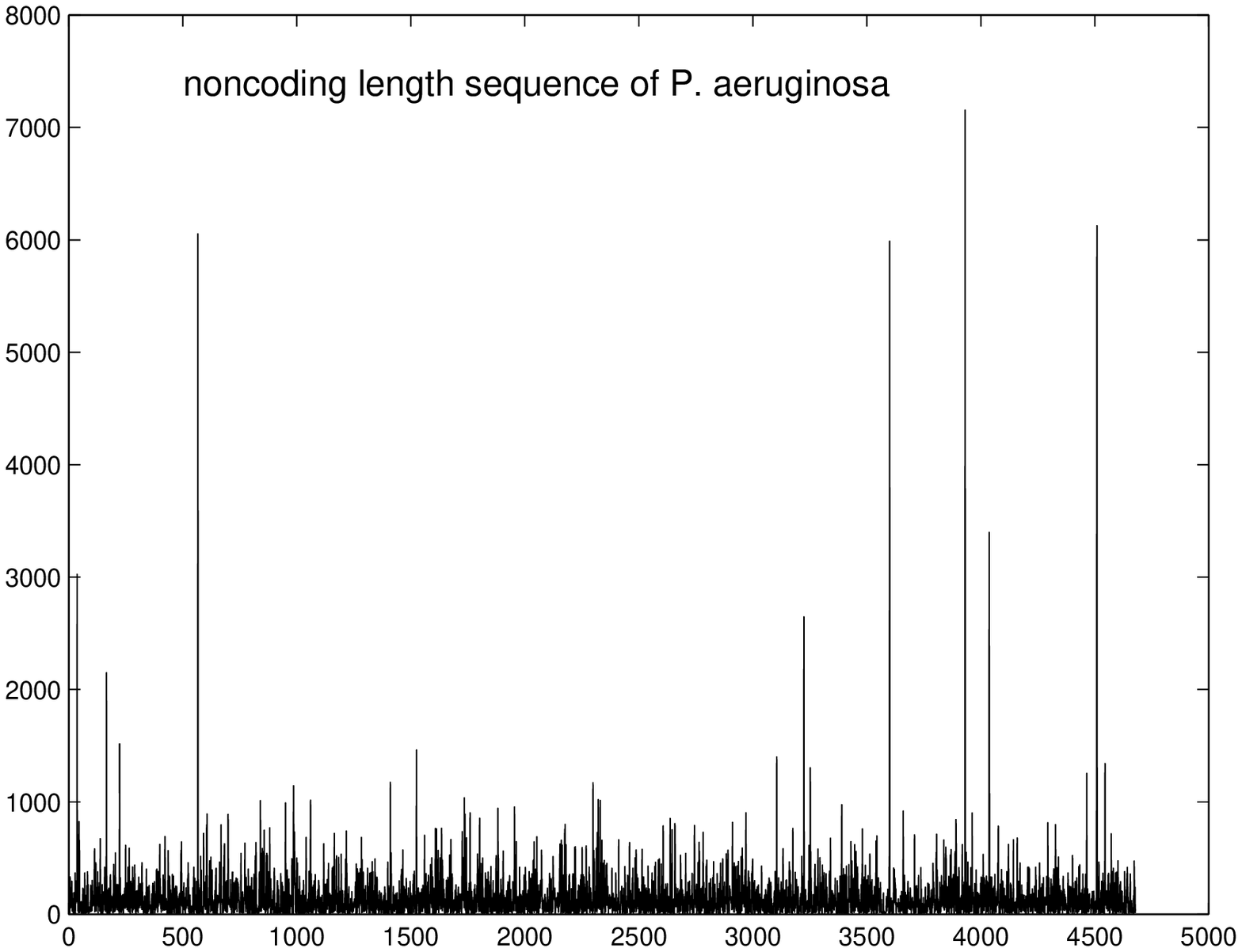}}
\caption{The coding and noncoding length sequences of {\it Pseudomonas
aeruginosa} }
\label{paerlength}
\end{figure}

First we counted out the length of coding and noncoding segments in the
complete genomes of the above bacteria and obtained the coding and noncoding
length sequences of these organisms. For example, we give the coding and
noncoding length sequences of {\it Pseudomonas aeruginosa} (paer) in Figure 
\ref{paerlength}.

\begin{figure}[tbp]
\centerline{\epsfxsize=8cm \epsfbox{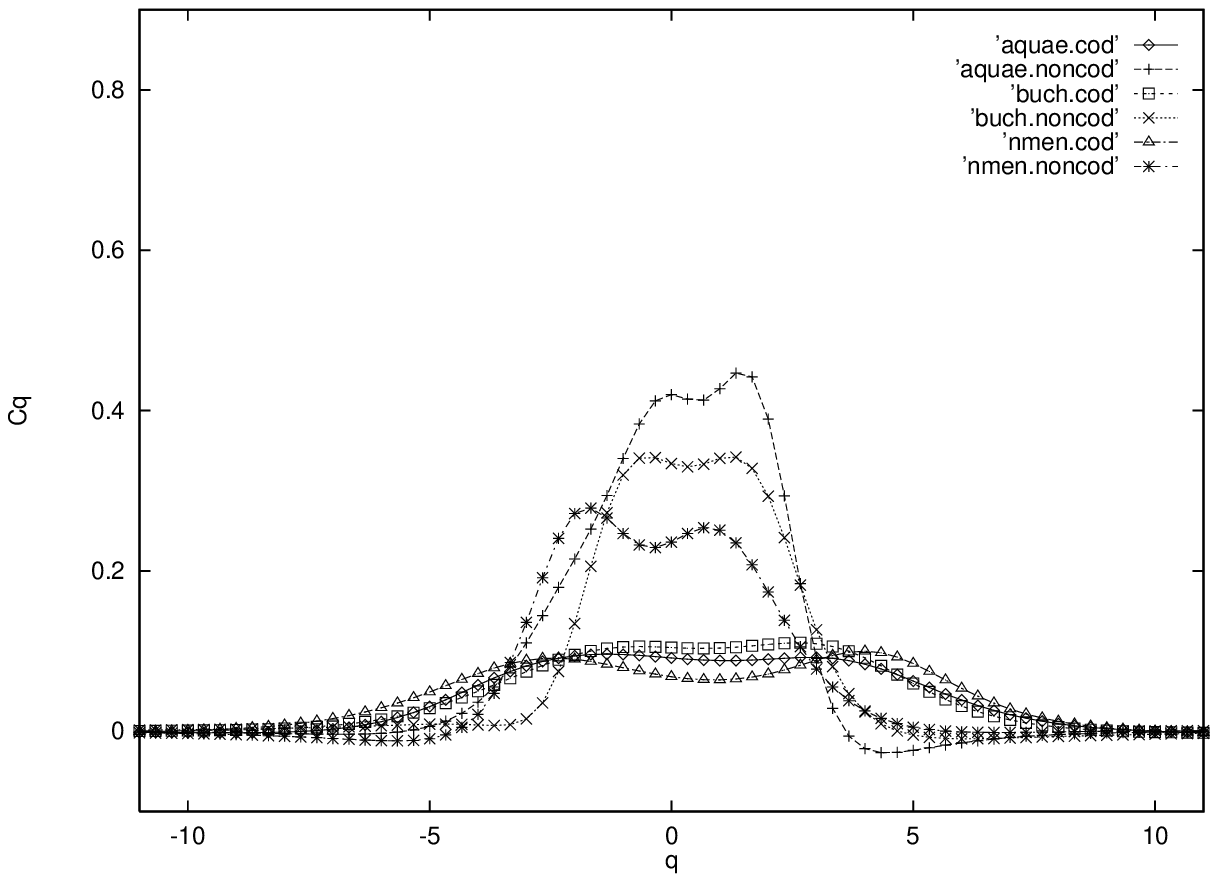}
\epsfxsize=8cm \epsfbox{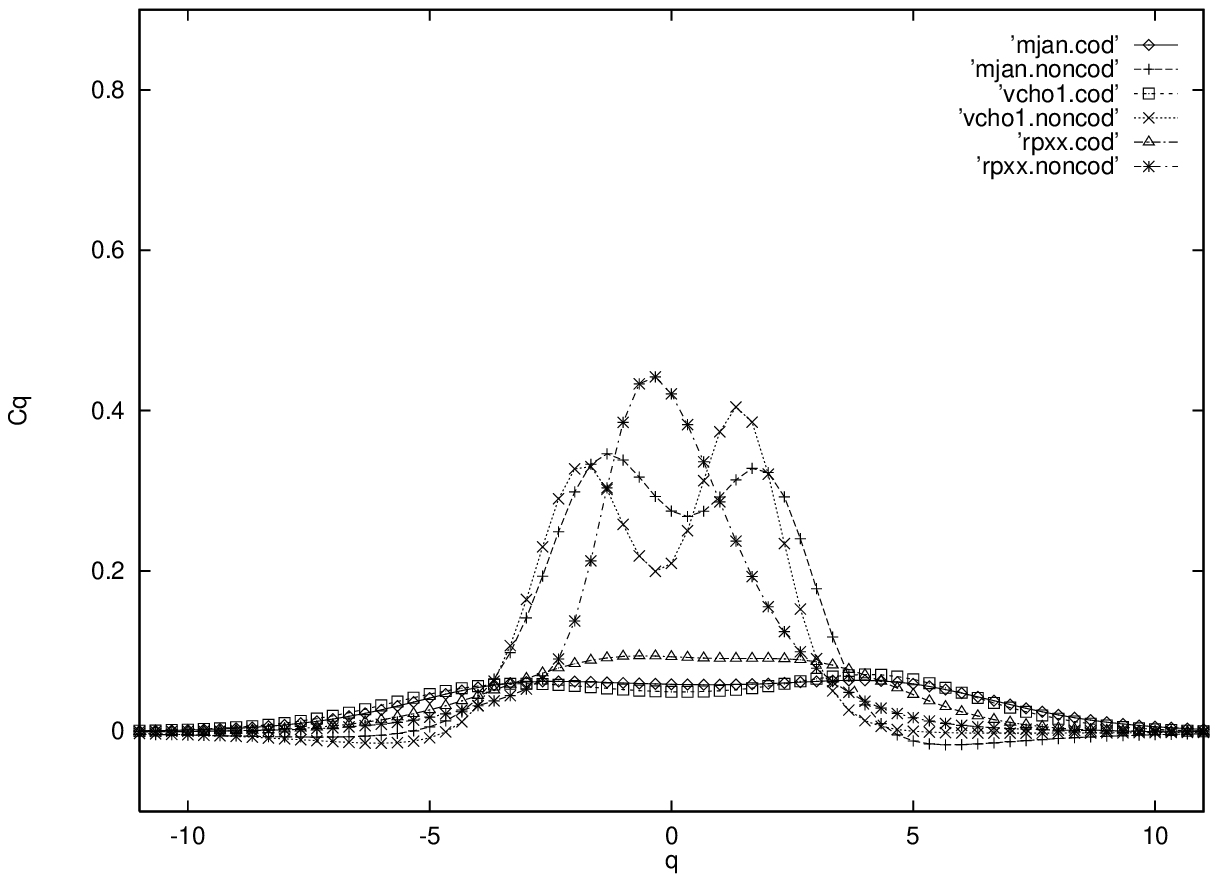}} 
\centerline{\epsfxsize=8cm \epsfbox{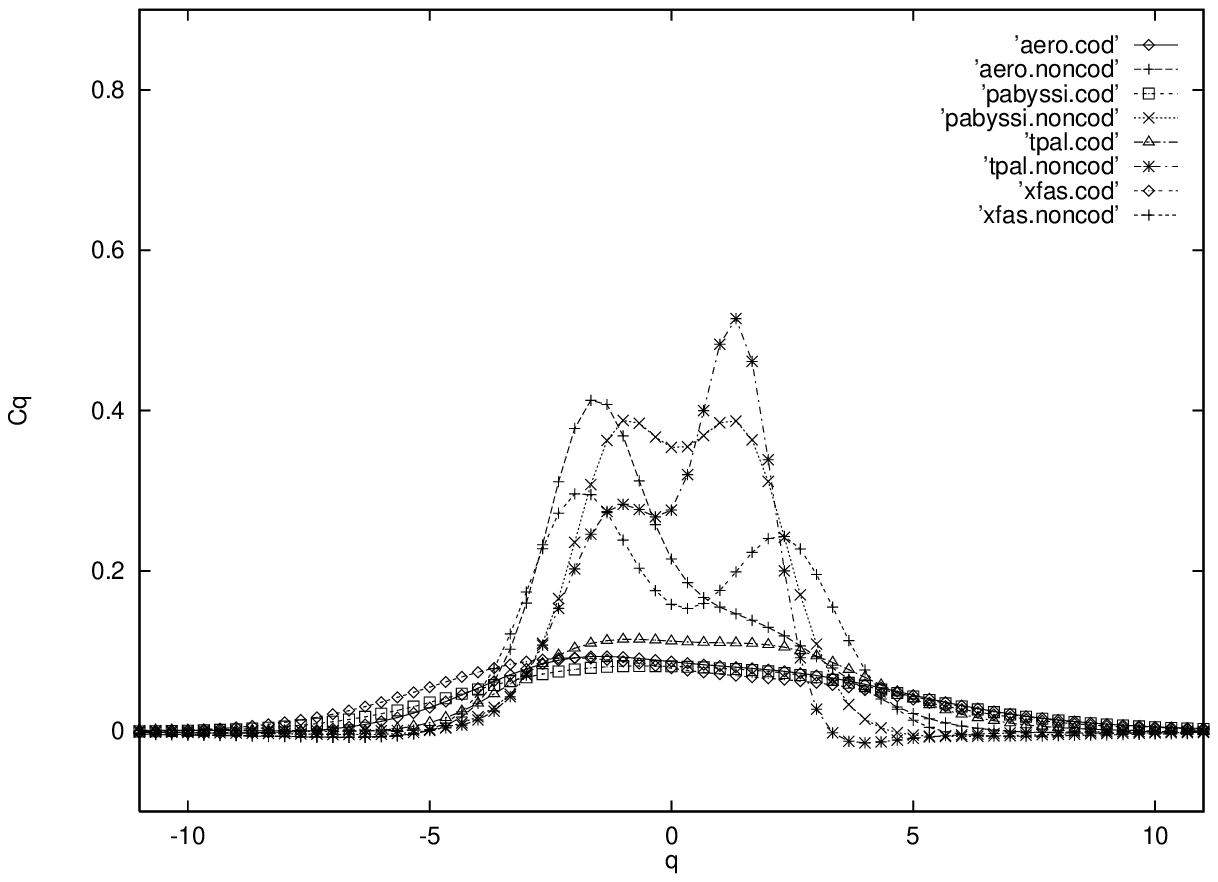}
\epsfxsize=8cm \epsfbox{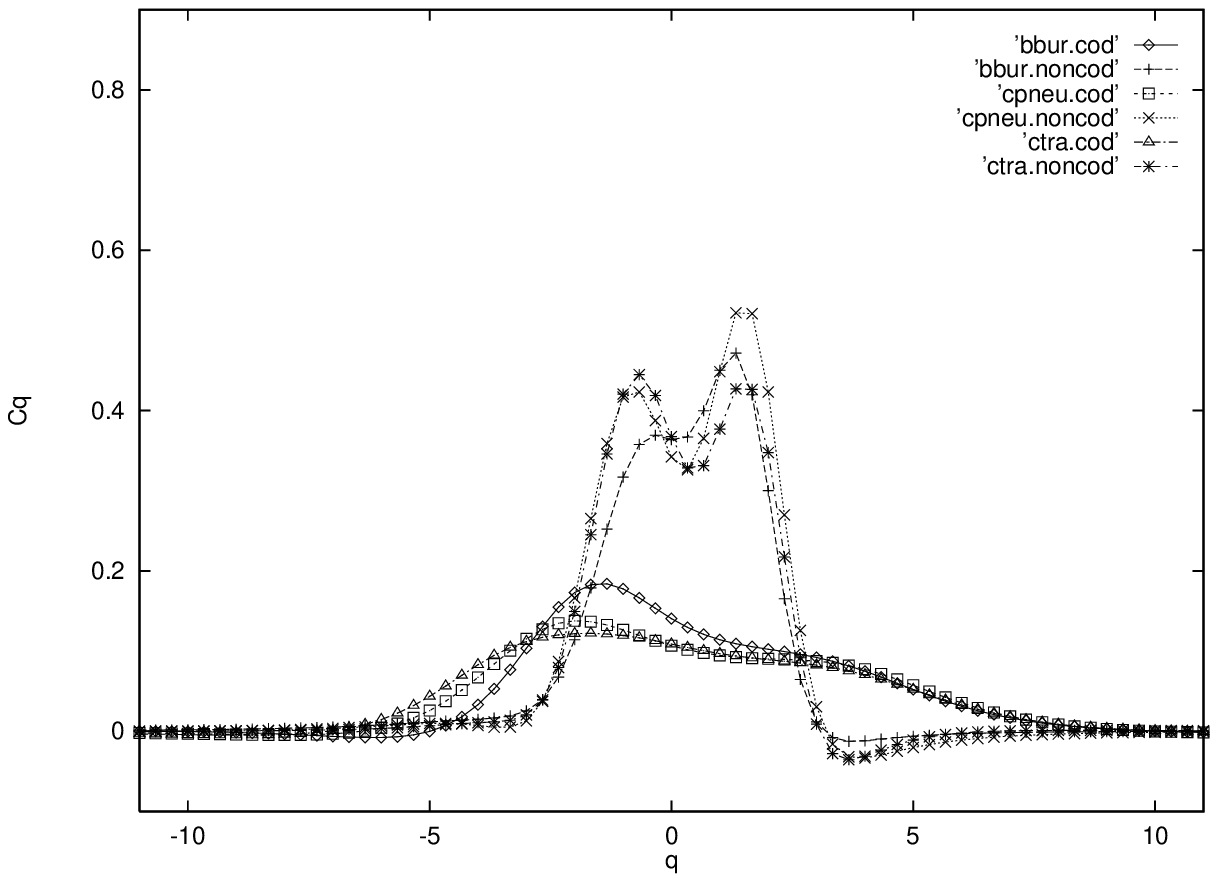}} 
\centerline{\epsfxsize=8cm \epsfbox{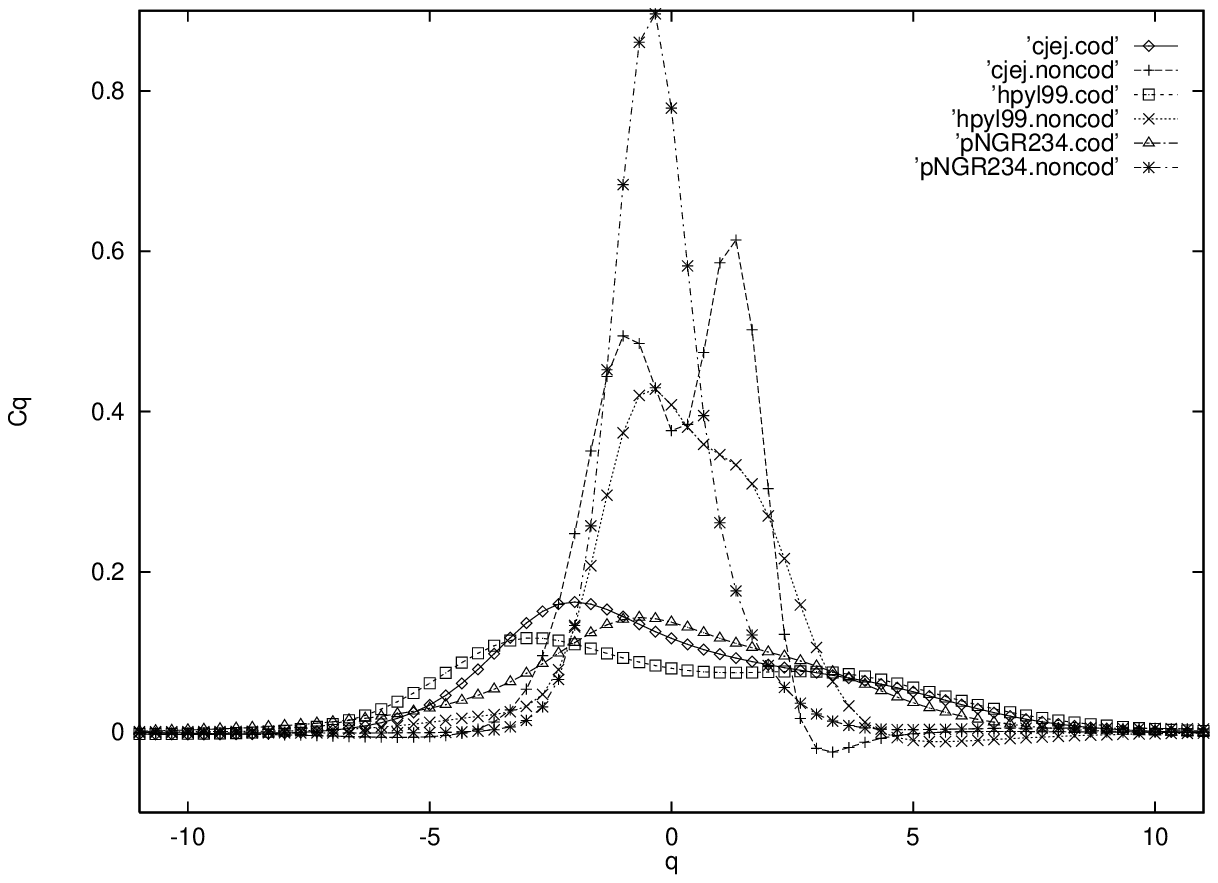}
\epsfxsize=8cm \epsfbox{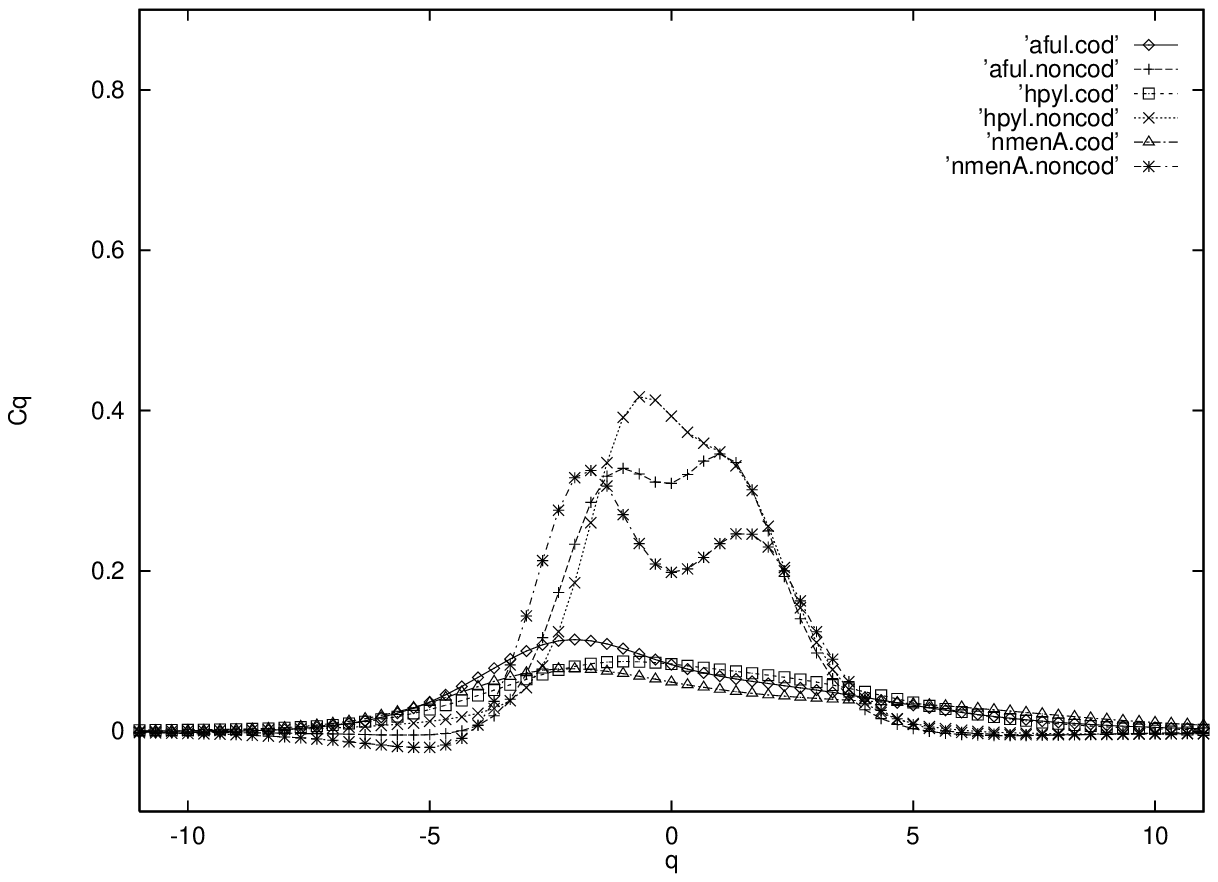}}
\caption{$C_q$ curves of coding and noncoding length sequences of 19
Bacteria. }
\label{Cqfigure1}
\end{figure}

\begin{figure}[tbp]
\centerline{\epsfxsize=8cm \epsfbox{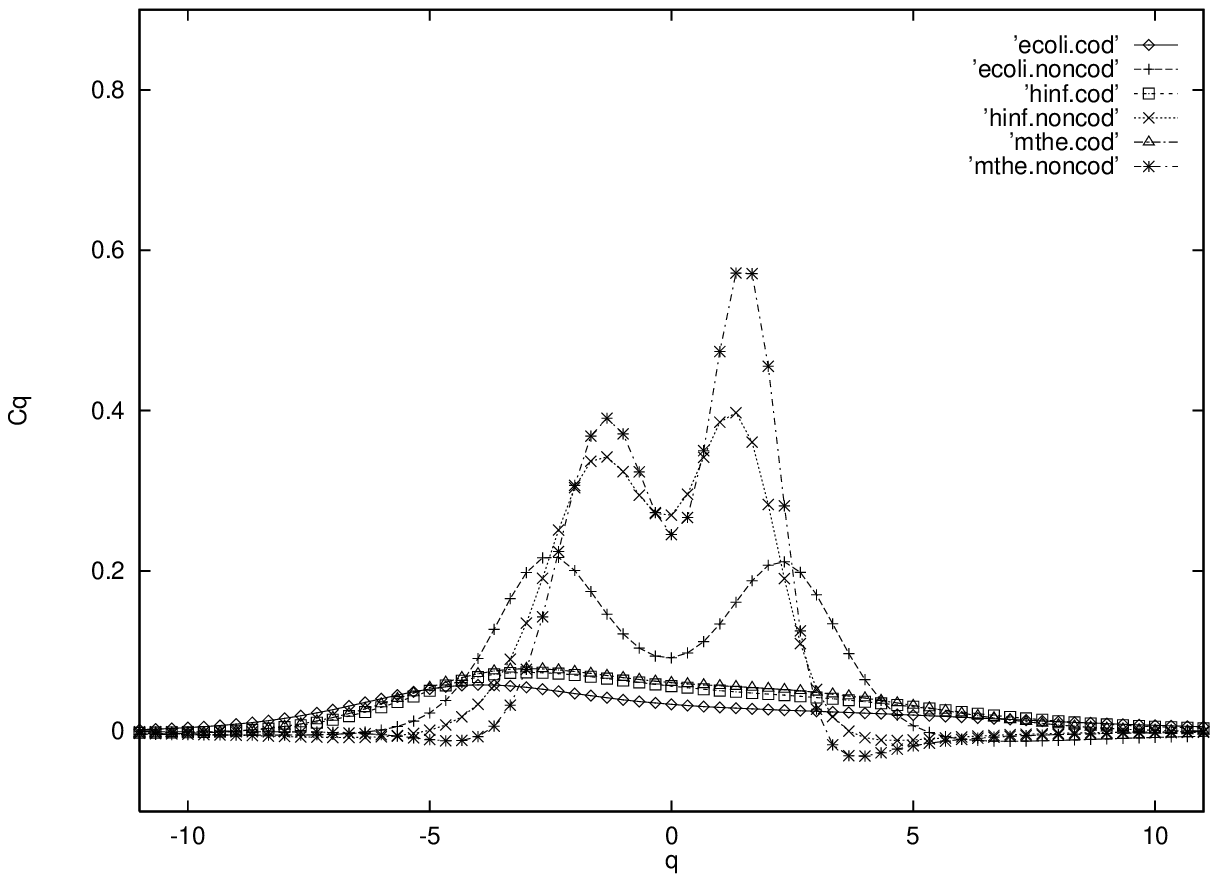}
\epsfxsize=8cm \epsfbox{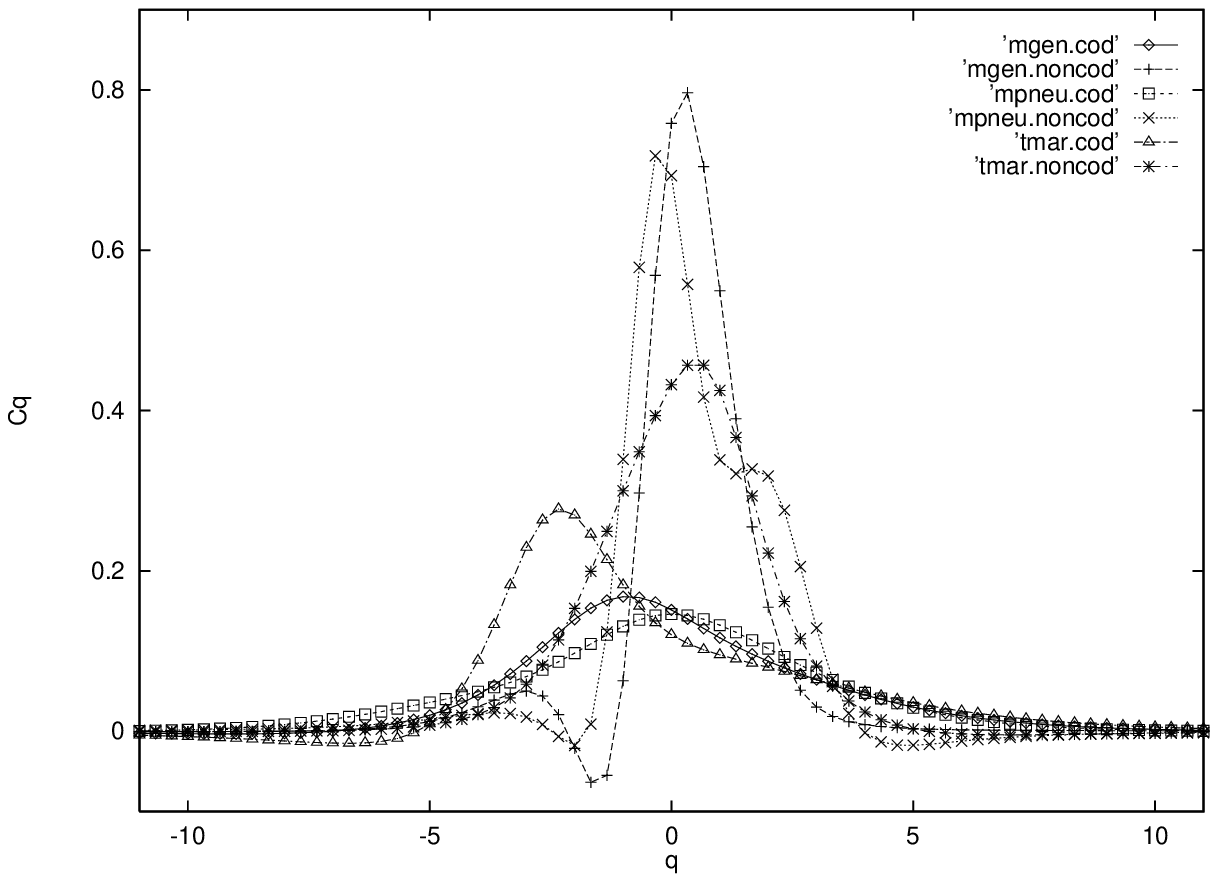}} 
\centerline{\epsfxsize=8cm \epsfbox{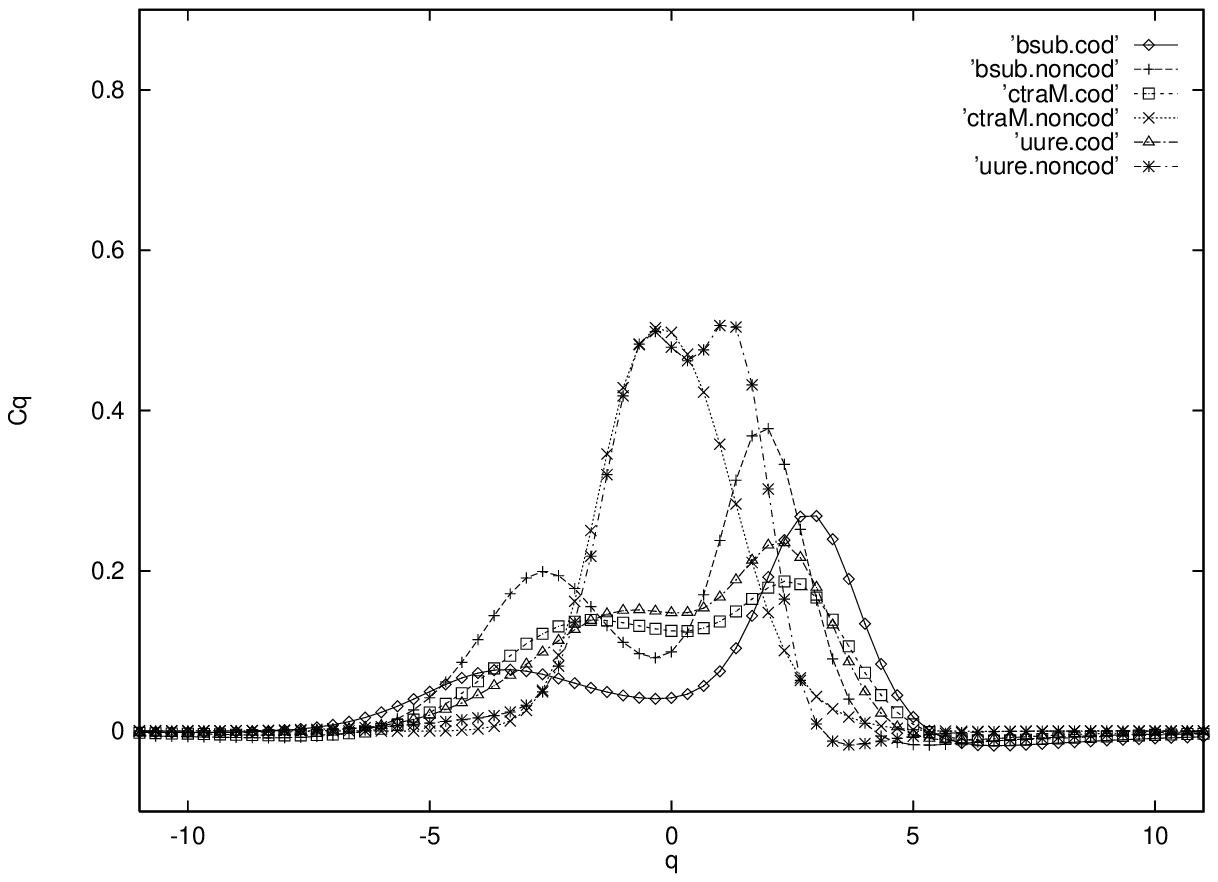}
\epsfxsize=8cm \epsfbox{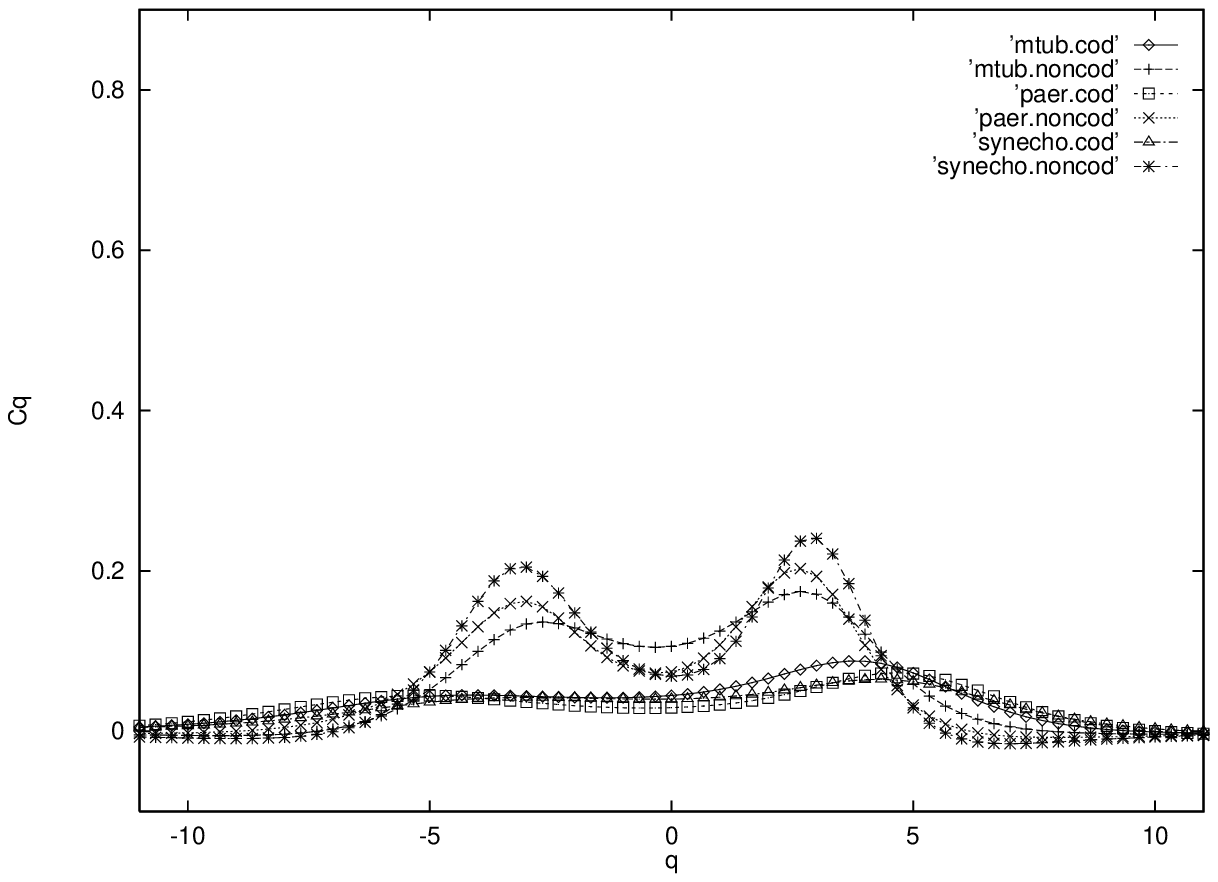}}
\caption{$C_q$ curves of coding and noncoding length sequences of another 12
Bacteria. }
\label{Cqfigure2}
\end{figure}

Then we calculated the dimension spectra $D_{q}$ and 'analogous' specific
heat $C_{q}$ of the coding and noncoding length sequences of all the above
bacteria according to the methods given in section 2. In order to show the
difference between coding and noncoding length sequences, we give the $C_{q}$
curves of length sequences of all the above bacteria as Figure \ref
{Cqfigure1} (for 19 bacteria) and Figure \ref{Cqfigure2} (for another 12
bacteria).

\begin{figure}[tbp]
\centerline{\epsfxsize=8cm \epsfbox{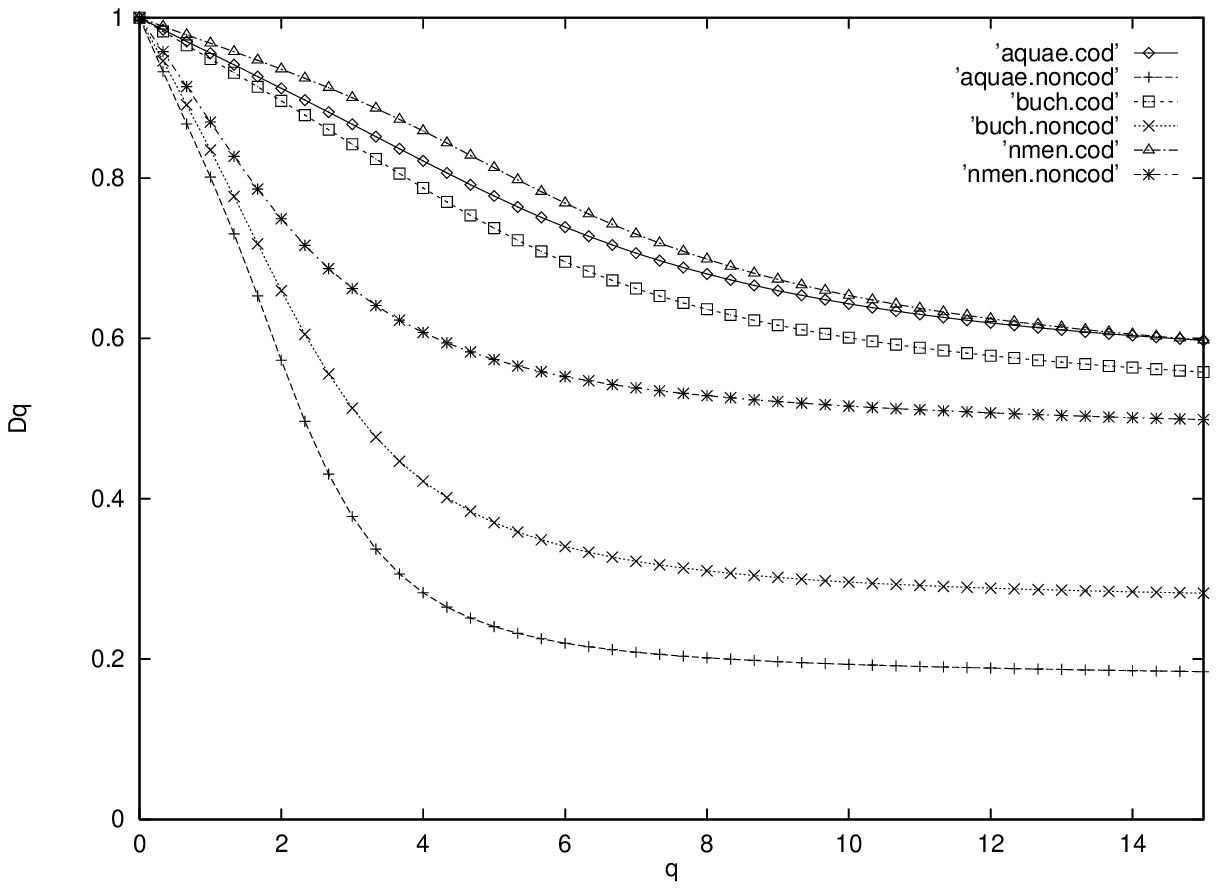}
\epsfxsize=8cm \epsfbox{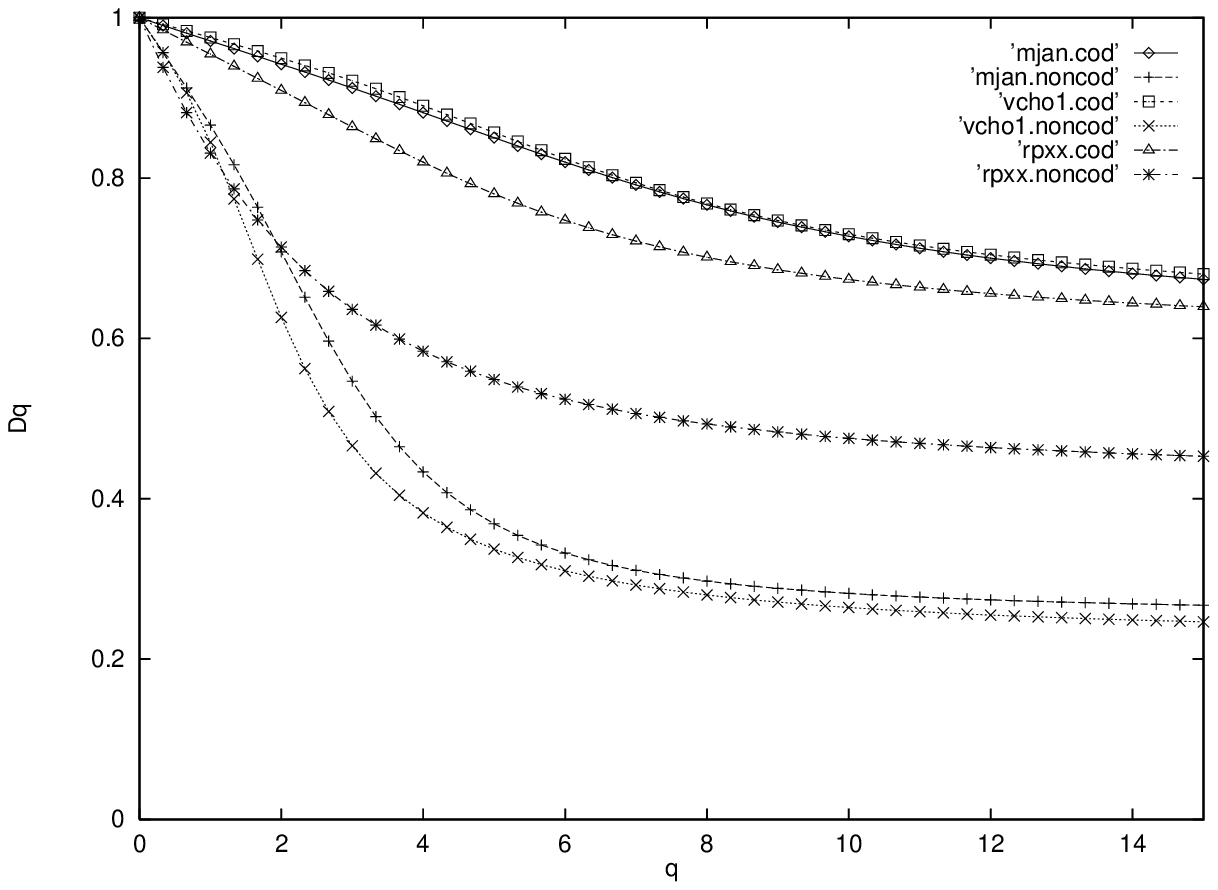}} 
\centerline{\epsfxsize=8cm \epsfbox{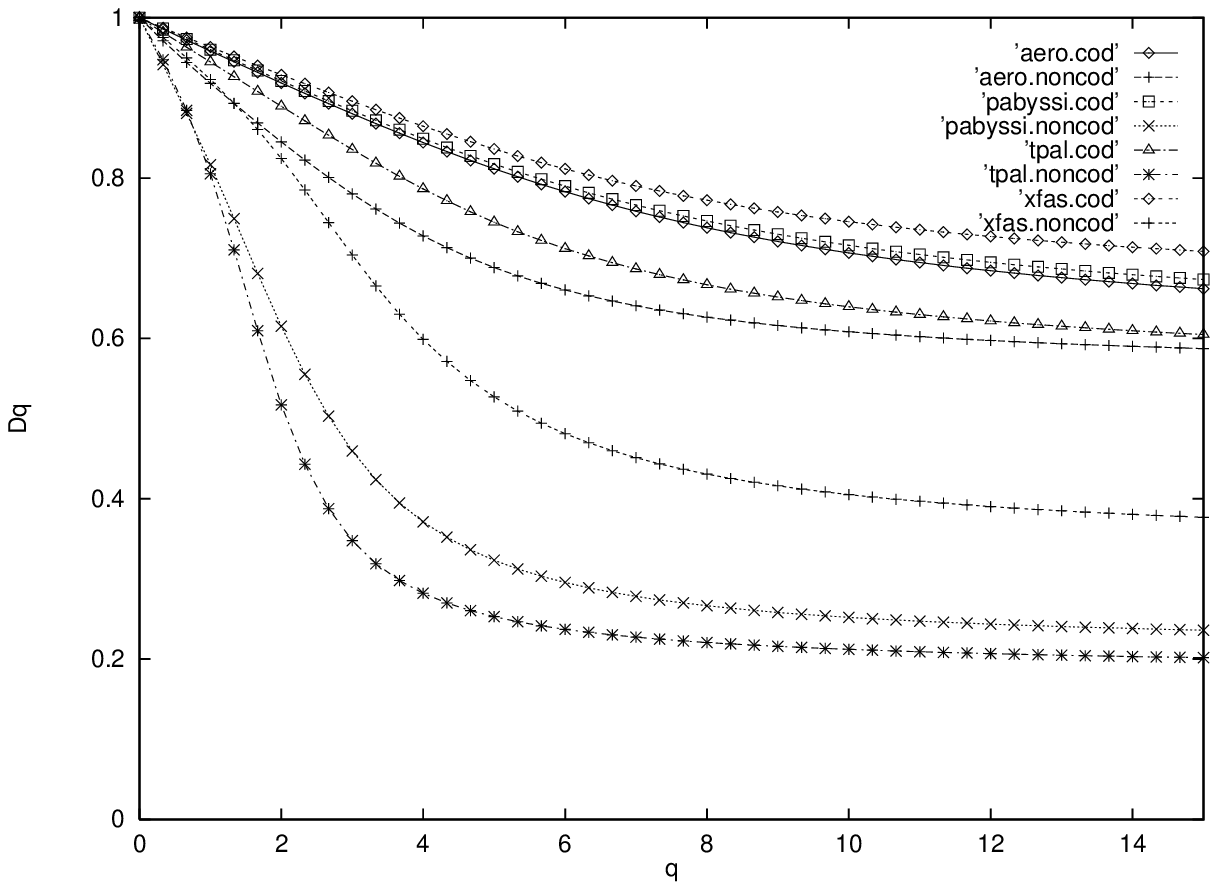}
\epsfxsize=8cm \epsfbox{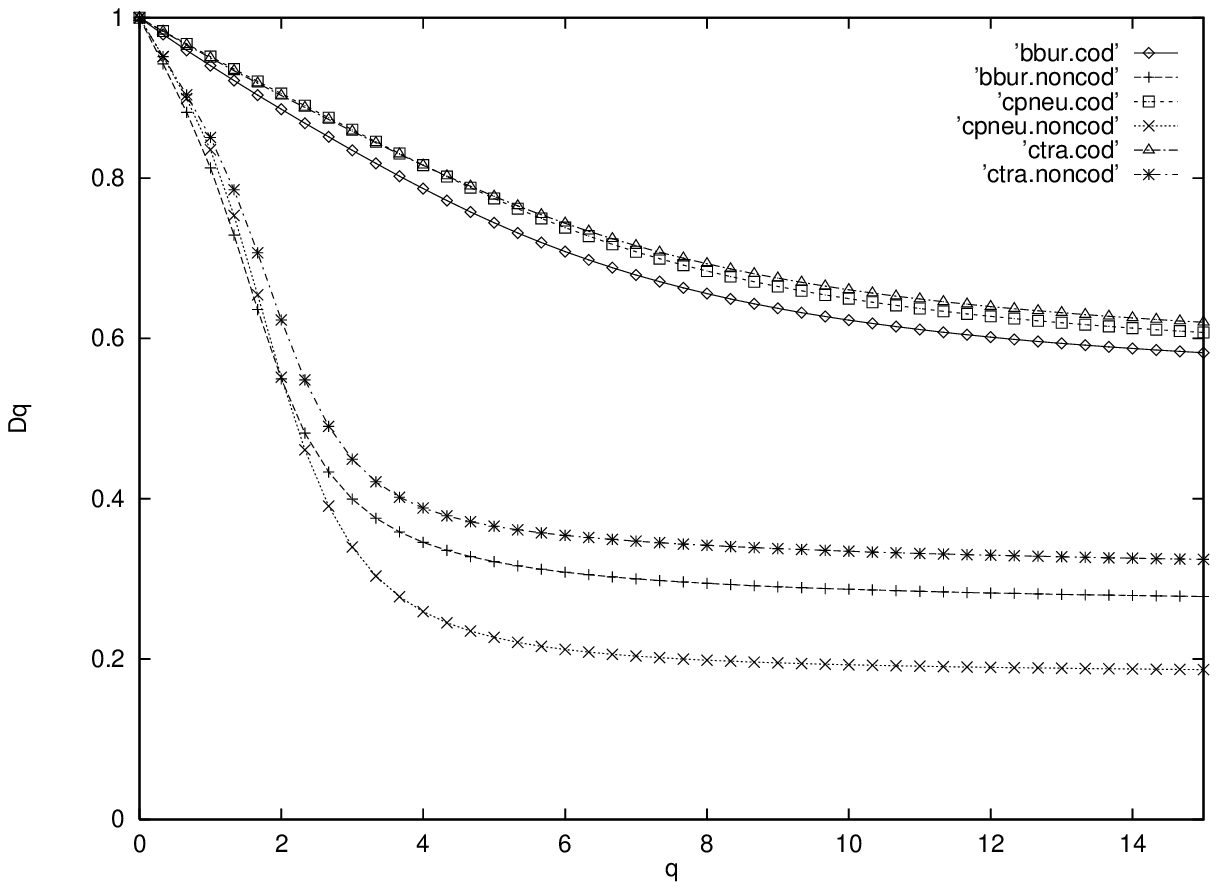}} 
\centerline{\epsfxsize=8cm \epsfbox{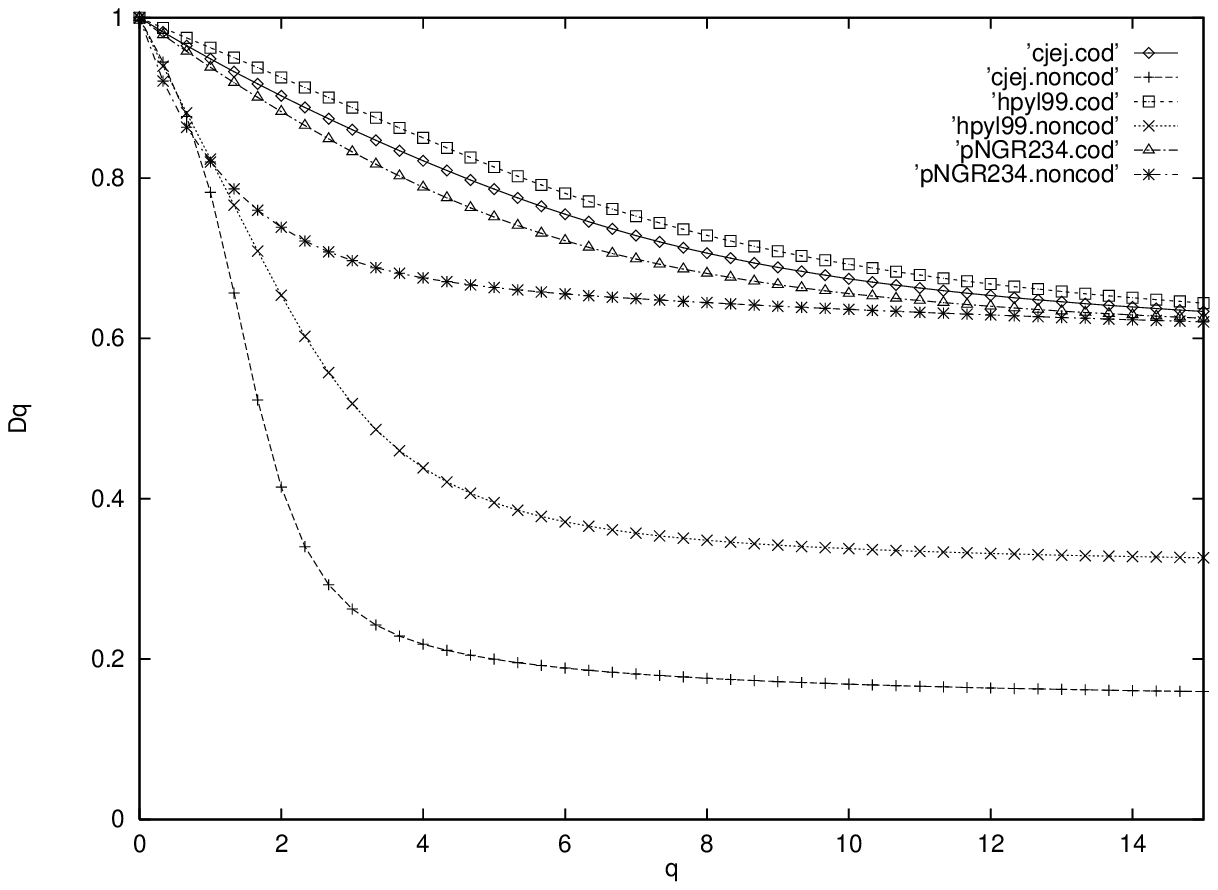}
\epsfxsize=8cm \epsfbox{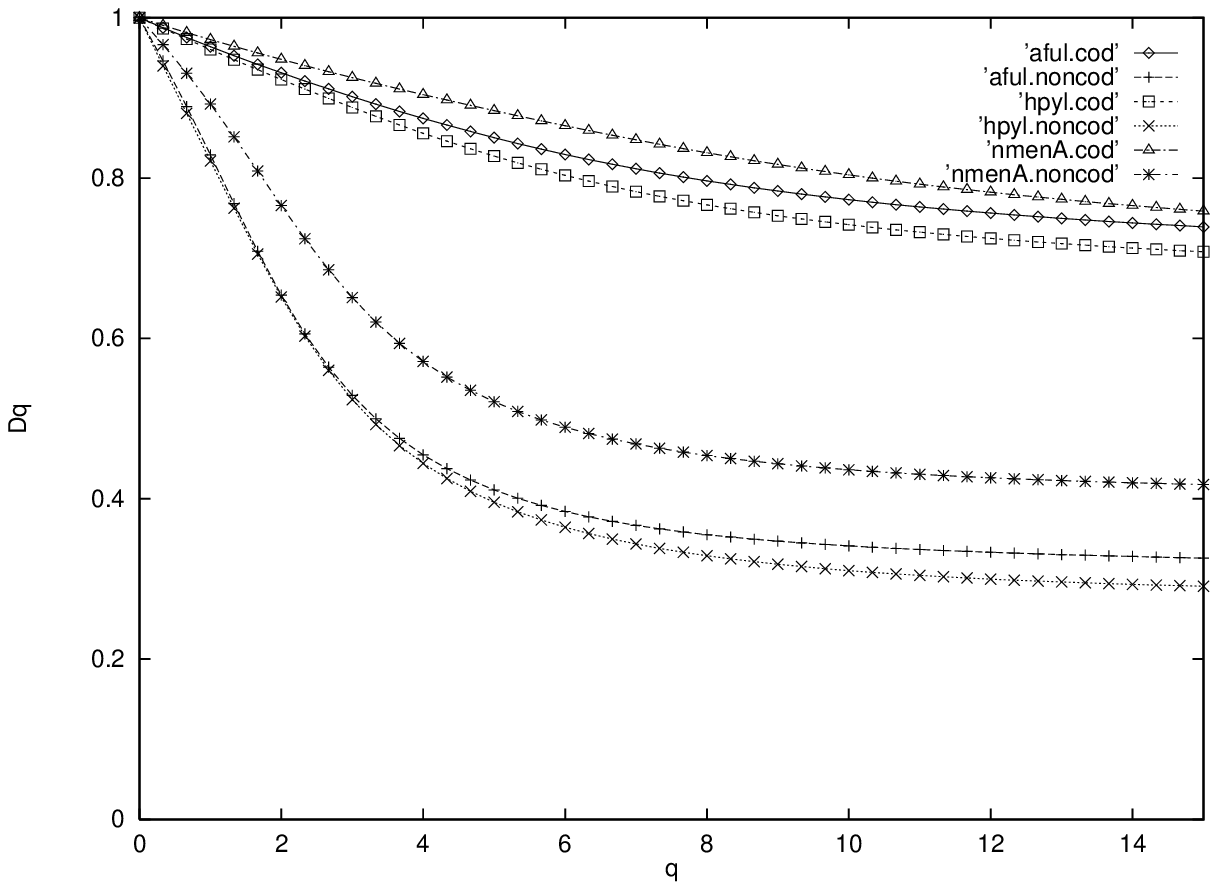}}
\caption{$D_q$ curves of coding and noncoding length sequences of 19
Bacteria. }
\label{Dqfigure1}
\end{figure}

\begin{figure}[tbp]
\centerline{\epsfxsize=8cm \epsfbox{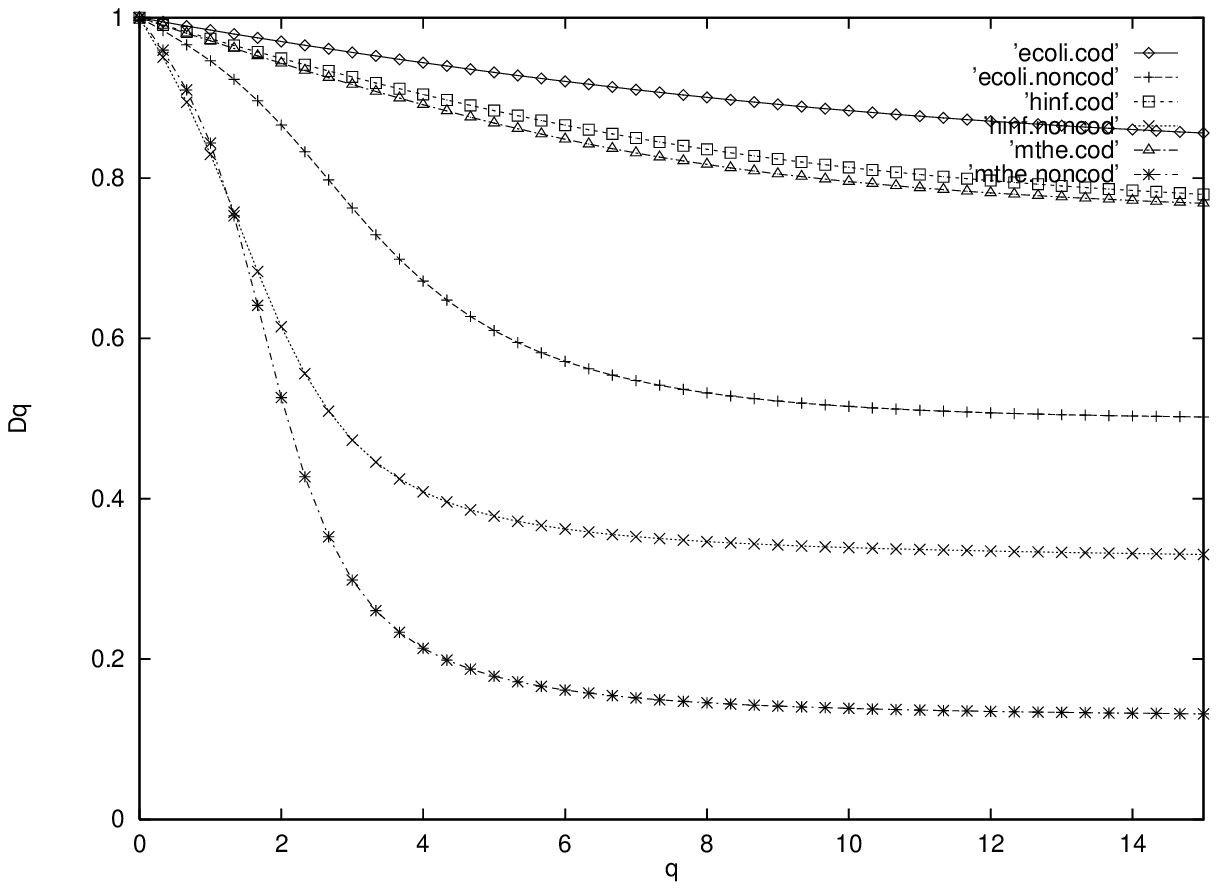}
\epsfxsize=8cm \epsfbox{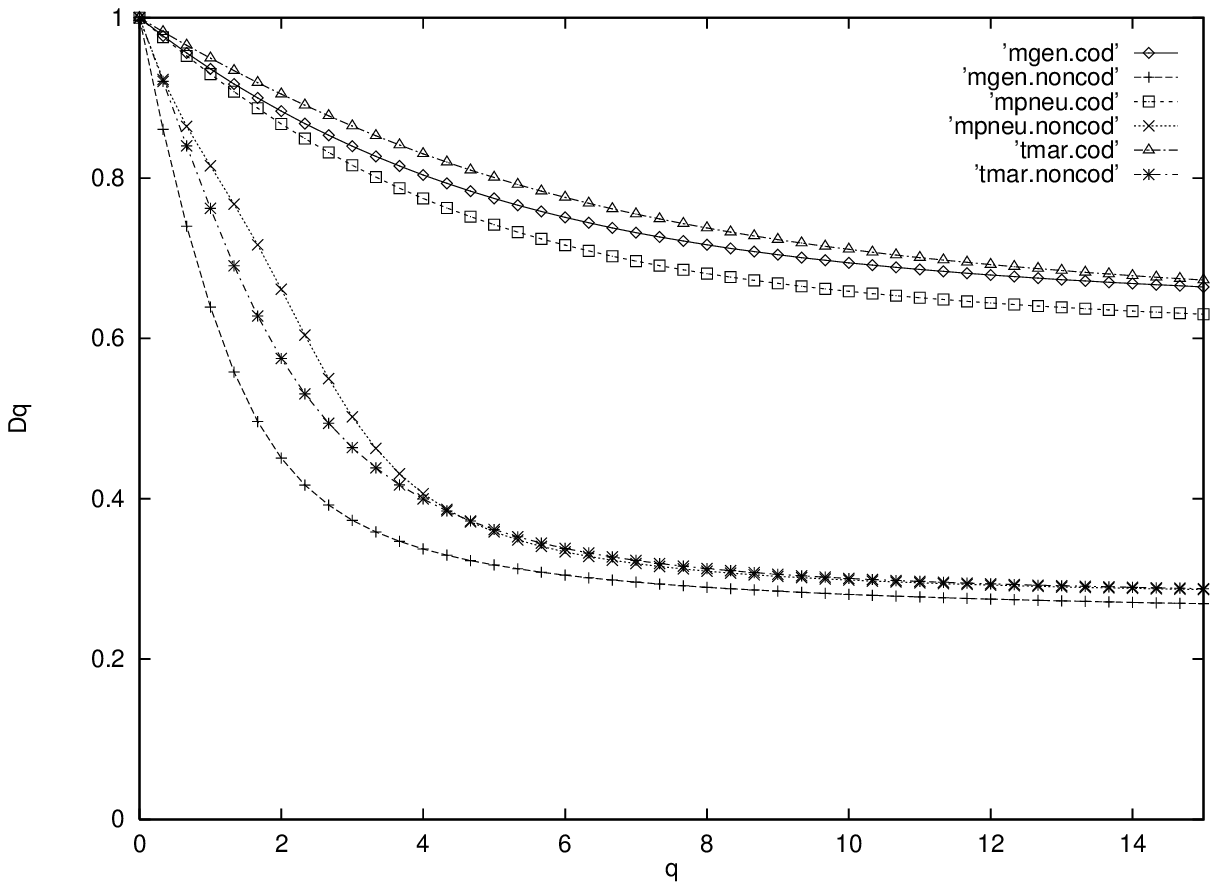}} 
\centerline{\epsfxsize=8cm \epsfbox{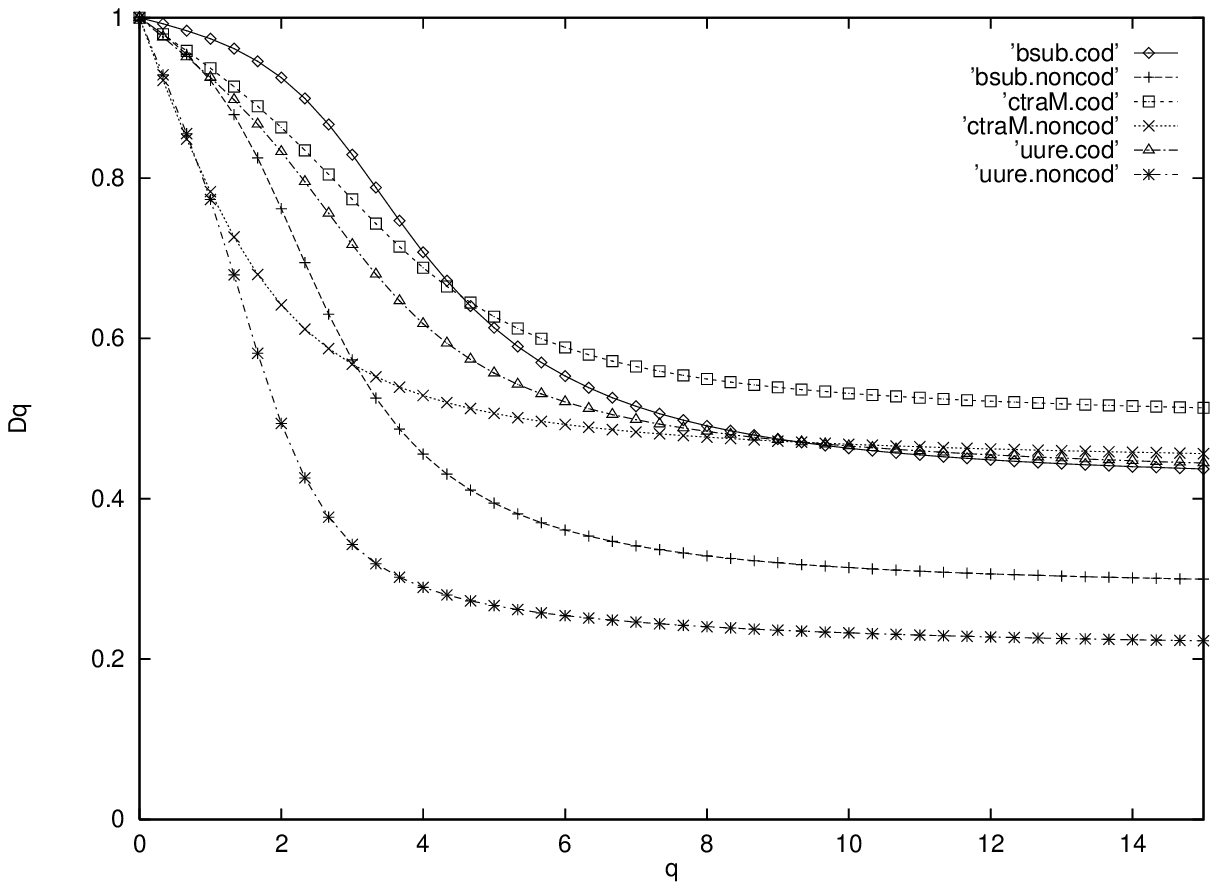}
\epsfxsize=8cm \epsfbox{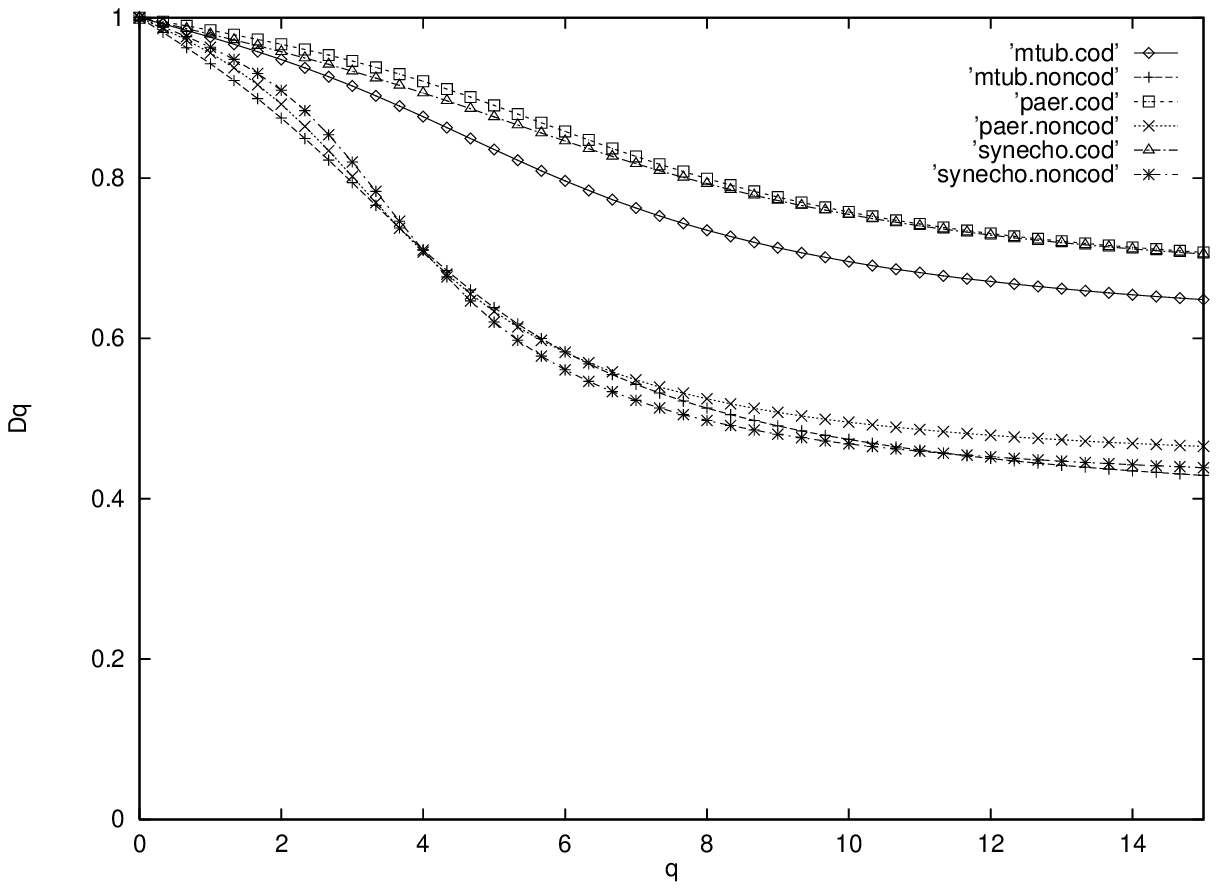}}
\caption{$D_q$ curves of coding and noncoding length sequences of another 12
Bacteria. }
\label{Dqfigure2}
\end{figure}

The hill behaviour of the dimension spectrum $D_{q}$ for $q<0$ is a well
known fact when using the box-counting method$^{\cite{pas97,can00}}$. In
Figures \ref{Dqfigure1} and \ref{Dqfigure2}, we present $D_{q}$ of the
coding or noncoding length sequences of all bacteria selected within the
range $q\geq 0$.

\section{Discussion and conclusions}

If a length sequence is completely random, then our measure definition
yields a uniform measure ($D_{q}=1,\ C_{q}=0$).

From the curves of $D_{q}$ and $C_{q}$, it is seen that there exists a clear
difference between the coding/noncoding length sequences of all organisms
considered here and the completely random sequence. Hence we can conclude
that complete genomes are not random sequences. But the $D_{q}$ values of
coding length sequences are closer to 1 than that of noncoding length
sequences. In other words, noncoding length sequences are further away from
a complete random sequence than coding length sequences. The property of the
length sequences is same as that of the DNA sequences$^{\cite{peng}}$.

We also found that for each bacterium selected, the $D_{q}$ values for $q>0$
of a noncoding length sequence are smaller than those of a coding length
sequence, but for $q<0$, the situation is reversed. It is well known that
the dimension is a measure for complexity. Here the complexity of noncoding
length sequences is higher than that of coding length sequences for bacteria.

From Figures \ref{Dqfigure1} and \ref{Dqfigure2}, almost all $D_{q}$ curves
for coding length sequences are flat, so their multifractality is not
pronounced. On the other hand, almost all $D_{q}$ curves for noncoding
length sequences are multifractal-like.

In our previous paper$^{\cite{YA00a}}$, we counted out all substrings with
fixed length appearing in the complete genome and gave a measure
representation of the complete genome. We found that the shape of the $C_{q}$
curves of all bacteria we selected are single-peaked. Hence this type of
phase transition of the measure representation is not useful for
classification of bacteria. On the other hand, from Figures \ref{Cqfigure1}
and \ref{Cqfigure2}, one can see that the 'analogous' specific heats of
noncoding length sequences of bacteria have a rich variety of behaviours
which is much more complex than that of coding length sequences. Some have
only one main single peak. In this class, some $C_{q}$ curves display a
shoulder to the right of the main peak, some display a shoulder to the left
of the main peak, and some have no shoulder, which resembles a classical
(first-order)phase transition at a critical point. In another class, the $%
C_{q}$ curves display a balance double-peak. So this provides a useful tool
for classification of bacteria according to the types of 'analogous'
specific heats of the noncoding length sequences. The relevant finding here
is that noncoding length sequences display higher $C_{q}$ peak heights and
clear double peaked structures than coding length sequences. This reveals
different types of long-range correlations between the two classes of
sequences. This new type of classification allows a better understanding of
the relationship among bacteria at the global gene level instead of
nucleotide sequence level. It can be useful to distinguish between sequence
curves as given in the example of Figure \ref{paerlength}.

To conclude, multifractal analysis provides a simple yet powerful method to
amplify the difference between a DNA length sequence and a random sequence.
In particular, the multifractal characterisation given by the 'analogous'
specific heat allows to distinguish DNA length sequences in more details.

\section*{ACKNOWLEDGEMENTS}

\ \ \ One of the authors, Zu-Guo Yu, would like to express his thanks to
Prof. Bai-lin Hao of Institute of Theoretical Physics of Chinese Academy of
Science for introducing him into this field and continuous encouragement.
The authors also thank Dr. Enrique Canessa for many good suggestions and
comments to improve this paper. The research was partially supported by
QUT's Postdoctoral Research Support Grant No. 9900658.

\newpage

\end{document}